\documentclass{article}

\usepackage{placeins}

\usepackage[utf8]{inputenc} 

\usepackage[T1]{fontenc}    
\usepackage{hyperref}       
\usepackage{url}            
\usepackage{booktabs}       
\usepackage{amsfonts}       
\usepackage{amsmath}        
\usepackage{nicefrac}       
\usepackage{microtype}      
\usepackage{lipsum}		
\usepackage{graphicx}
\usepackage{natbib}
\usepackage{doi}

\usepackage{arxiv}

\renewcommand{\headeright}{DUAL-ATTENTION RESNET OUTPERFORMS TRANSFORMERS IN HER2 PREDICTION ON DCE-MRI}
\renewcommand{\undertitle}{Preprint}
\renewcommand{\shorttitle}{}

\hypersetup{
pdftitle={DUAL-ATTENTION RESNET OUTPERFORMS TRANSFORMERS IN HER2 PREDICTION ON DCE-MRI},
pdfsubject={Medical Imaging, Deep Learning, Breast Cancer},
pdfauthor={Naomi Fridman, Anat Goldstein},
pdfkeywords={HER2 prediction, DCE-MRI, deep learning, attention mechanisms, breast cancer}
}

\title{Dual-Attention ResNet outperform transformers in HER2 prediction on DCE-MRI}

\author{
Naomi Fridman$^{1,*}$ and Anat Goldstein$^{1}$\\
$^{1}$Department of Industrial Engineering, Ariel University, Ariel 40700, Israel\\
$^{*}$Corresponding author: naominoe.fridman@msmail.ariel.ac.il
}

\begin{document}
\maketitle

\begin{abstract}
Breast cancer is the most diagnosed cancer in women, with human epidermal growth factor receptor 2 (HER2) status critically guiding treatment selection. Reliable, noninvasive prediction of HER2 status directly from dynamic contrast-enhanced magnetic resonance imaging (DCE-MRI) could streamline diagnostics and reduce reliance on invasive biopsy. However, preprocessing high-dynamic-range (12--16 bit) DCE-MRI into standardized 8-bit RGB format for pretrained neural networks is nontrivial; normalization strategy can heavily impact downstream model accuracy.

We systematically benchmarked intensity normalization strategies using a
Triple-Head Dual-Attention ResNet designed to process RGB-fused temporal sequences
from three DCE phases (pre-contrast, early post-contrast, and late post-contrast). The model was trained and validated on a multicenter cohort (n=1,149) from the Investigation of Serial Studies to Predict Your Therapeutic Response with Imaging and Molecular Analysis (I-SPY) trials, with external validation on the independent BreastDCEDL\_AMBL dataset (n=43 lesions). Under matched preprocessing and training protocols, the attention-based
ResNet outperformed transformer-based models (Convolutional Vision Transformer, Vision
Transformer), achieving an accuracy of 0.75 and an area under the Receiver Operating Characteristic curve (AUC) of
0.74 on I-SPY test data. N4 bias field correction, widely used in radiomics, slightly degraded deep
learning performance. Without fine-tuning, cross-institutional external validation achieved an
AUC of 0.66, demonstrating generalizability across imaging protocols.

These results indicate that dual-attention mechanisms effectively capture
transferable spatiotemporal features in DCE-MRI for HER2 stratification, advancing
reproducible and generalizable deep learning biomarkers for breast cancer.
\end{abstract}



\section{Introduction}

Breast cancer is the most frequently diagnosed cancer among women worldwide, accounting for approximately 30\% of all female cancers~\cite{turnbull2009dynamic}. DCE-MRI has become an essential tool in breast cancer diagnosis and management, providing superior soft tissue contrast and functional information through temporal visualization of contrast agent kinetics. DCE-MRI captures both morphological features and physiological processes including tumor angiogenesis, vascular permeability, and perfusion patterns that reflect underlying tumor biology~\cite{kuhl2007current}.

Molecular stratification of breast cancer is guided by assessment of estrogen receptor (ER), progesterone receptor (PgR), and HER2 status, defining clinically actionable subtypes~\cite{blows2010subtyping,weigelt2010gene,cheang2009ki67,vallejos2010classification}. Clinicopathologic factors such as tumor size, grade, nodal involvement, histologic type, and margins further inform prognosis and individualized therapy~\cite{cheang2009ki67,vallejos2010classification}. HER2 overexpression, present in approximately 15--20\% of breast cancers, enables effective targeted therapies ~\cite{slamon2001use, wolff2018her2} but standard assessment via biopsy followed by immunohistochemistry or fluorescence in situ hybridization is invasive, labor-intensive, and subject to spatial sampling bias and inter-observer variability. A reliable imaging-based method for HER2 prediction from DCE-MRI could improve diagnostic workflow, reduce patient burden, and enable earlier treatment planning.

While some radiomics and deep learning studies have shown promise, most HER2 prediction models are trained and evaluated on small, single-center MRI datasets, frequently using handcrafted features and lacking demonstration of generalizability~\cite{luo2024radiomics,fan2025her2}. Models such as CNNs\cite{thatha2025deep,behar2021resnet50} and, more recently,
Vision Transformers (ViT) and Convolutional Vision Transformers
(CvT)\cite{dosovitskiy2020transformers,fridman2025breastdcedl,wang2017residual,wei2022mosquito},
have achieved strong results in medical image classification.classification. Yet, a comparative evaluation of these modern architectures for HER2 prediction from DCE-MRI in a multicenter setting has not been systematically conducted.

Dual-attention ResNet architectures, which incorporate both channel and spatial attention mechanisms, have demonstrated performance gains for diverse clinical imaging tasks including histopathology classification, denoising, and image quality assessment~\cite{liu2023dalaresnet,hu2024csran,zhu2024ctquality,xu2023resnetreview,cheng2022resganet}. Attention modules allow models to adaptively focus on the most informative features and spatial regions, which is especially valuable in heterogeneous medical images.

A critical yet often overlooked challenge in applying deep learning to DCE-MRI
is preprocessing, particularly the conversion of high-dynamic-range Digital Imaging and
Communications in Medicine data (typically 12--16 bit, with voxel intensities exceeding
3,000--12,000) into the standardized 8-bit RGB format (0--255) required by pretrained
ImageNet backbones. This massive data reduction can dramatically impact model performance depending on the normalization strategy employed. Our study serves as a systematic benchmark for preprocessing methods in DCE-MRI analysis, evaluating seven distinct normalization and clipping strategies. Notably, we found
that N4 bias field correction—commonly used in radiomics pipelines to correct intensity inhomogeneities—slightly
degraded deep learning performance in our experiments. This finding suggests that for attention-based deep learning, raw intensity distributions may preserve diagnostically relevant patterns that bias correction inadvertently removes, representing a significant time-saving opportunity by eliminating this computationally expensive preprocessing step.

To enable robust benchmarking, we assembled models on the BreastDCEDL dataset~\cite{fridman2025breastdcedl,breastdcedlds}, comprising 1,149 pre-treatment 3D DCE-MRI scans from the Investigation of Serial Studies to Predict Your Therapeutic Response with Imaging and Molecular Analysis (I-SPY) trials~\cite{spy1,spy2a,spy2b}, harmonized and converted to standardized NIfTI volumes with unified annotations. This multicenter dataset enables systematic comparison of state-of-the-art architectures (ViT, CvT, Triple-Head Dual-Attention ResNet) for HER2 prediction using consistent protocols.

For external validation and to examine model generalizability, we evaluated our trained models without fine-tuning on the independent, single-center BreastDCEDL\_AMBL dataset (n=43 lesions), which differs by vendor, protocol, and patient population. Our Triple-Head Dual-Attention ResNet achieved Area Under the Receiver
Operating Characteristic Curve (AUC) 0.74 on I-SPY test data and maintained and maintained
reasonable discriminative performance (AUC 0.61--0.66) on AMBL despite substantial domain
shift, indicating that attention-driven architectures capture transferable imaging features
even without retraining.

Overall, this work leverages a multicenter, standardized dataset to systematically compare preprocessing strategies and leading deep learning architectures for DCE-MRI-based HER2 prediction, providing evidence for cross-institutional generalizability and practical insights for clinical implementation.

\section{Methods}

\subsection{Study Population}

The study cohort included 1,149 breast cancer patients from the BreastDCEDL ~\cite{fridman2025breastdcedl}, namely  I-SPY clinical trials~\cite{spy2a,spy2b,spy1}. Participants were split to training (n=885, 77.0\%), testing (n=132, 11.5\%), and validation (n=132, 11.5\%) subgroups, following the BreastDCEDL defined benchmark. Distributions of age, hormone receptor (HR) and HER2 status, race, and pathologic complete response (pCR) were well-balanced across splits (Table~\ref{tab:ispy_demo}), supporting robust model development and unbiased evaluation.

\begin{table}[ht]
\centering
\caption{I-SPY Trial Patient Demographics and Clinical Characteristics (n=150) at
baseline.}
\label{tab:ispy_demo}
\begin{tabular}{lcccc}
\hline
Characteristic & Training & Testing & Validation & Total \\
\hline
Number of patients & 885 & 132 & 132 & 1,149 \\
Age, mean $\pm$ SD (years) & 50.1 $\pm$ 10.5 & 50.5 $\pm$ 10.5 & 50.4 $\pm$ 10.8 & 50.2 $\pm$ 10.5 \\
\textbf{Race, n (\%)} & & & & \\
\quad White             & 662 (74.8) & 96 (72.7) & 96 (72.7) & 854 (74.3) \\
\quad Black             & 139 (15.7) & 25 (18.9) & 25 (18.9) & 189 (16.4) \\
\quad Other/Unknown     & 84 (9.5)   & 11 (8.3)  & 11 (8.3)  & 106 (9.2) \\
\textbf{Biomarkers/Response, n (\%)} & & & & \\
\quad HR-positive       & 556 (62.8) & 86 (65.2) & 85 (64.4) & 727 (63.3) \\
\quad HER2-positive     & 213 (24.1) & 32 (24.2) & 34 (25.8) & 279 (24.3) \\
pCR achieved           & 234 (35.4) & 29 (29.0) & 29 (29.0) & 292 (34.3) \\
\hline
\end{tabular}
\\[1ex]
\begin{minipage}{0.97\linewidth}
\small Data are presented as n (\%) unless otherwise specified. HR, hormone receptor; HER2, human epidermal growth factor receptor 2; pCR, pathologic complete response; SD, standard deviation.
\end{minipage}
\end{table}

\subsection{DCE-MRI Preprocessing and Normalization}

Preprocessing of DCE-MRI data is critical for reliable deep learning, as DICOM-encoded intensities commonly present long-tailed distributions. Typical voxel values often exceed 3,000–12,000 units. However, pretrained ImageNet backbones require input mapped to the standard 8-bit range (0–255). The normalization strategy can significantly affect contrast preservation and downstream model performance~\cite{kubassova2007quantitative}.

We evaluated seven normalization strategies for mapping multicenter DCE-MRI scans into a consistent 0–255 range.

\paragraph{Per-slice min-max normalization.} Each slice is normalized independently:
\begin{equation}
x' = \left( \frac{x - \min(x_{\text{slice}})}{\max(x_{\text{slice}}) - \min(x_{\text{slice}})} \right) \times 255.
\label{eq:slice_norm}
\end{equation}

\paragraph{Global min-max normalization.} Extremal values are computed across the entire 3D volume:
\begin{equation}
x' = \left( \frac{x - \min(x_{\text{volume}})}{\max(x_{\text{volume}}) - \min(x_{\text{volume}})} \right) \times 255.
\label{eq:global_norm}
\end{equation}

\paragraph{Lower quantile clipping.} The 10th percentile is used as the lower bound:
\begin{equation}
x' = \left( \frac{\max(x - q_{0.1}, 0)}{\max(x_{\text{volume}}) - q_{0.1}} \right) \times 255.
\label{eq:lower_clip}
\end{equation}

\paragraph{Channelwise upper clipping.} Each temporal phase $t$ is clipped at its 98th percentile:
\begin{equation}
C_{98}^{(t)} = \text{quantile}_{0.98}(x^{(t)}).
\label{eq:clip_threshold}
\end{equation}
\begin{equation}
x'^{(t)} = \min\left( \frac{x^{(t)}}{C_{98}^{(t)}}, 1 \right) \times 255.
\label{eq:channel_clip}
\end{equation}

\paragraph{Global upper clipping.} A single threshold is computed across all temporal phases:
\begin{equation}
x' = \min\left( \frac{x}{C_{98}^{\text{global}}}, 1 \right) \times 255.
\label{eq:global_clip}
\end{equation}

\paragraph{N4 bias field correction.} Each scan is optionally corrected using the N4ITK algorithm~\cite{tustison2010n4itk}, which estimates and removes low-frequency intensity inhomogeneities. This step improves inter-scan comparability but may degrade performance in attention-based models.

\begin{figure}[htbp]
\centering
\includegraphics[width=\linewidth]{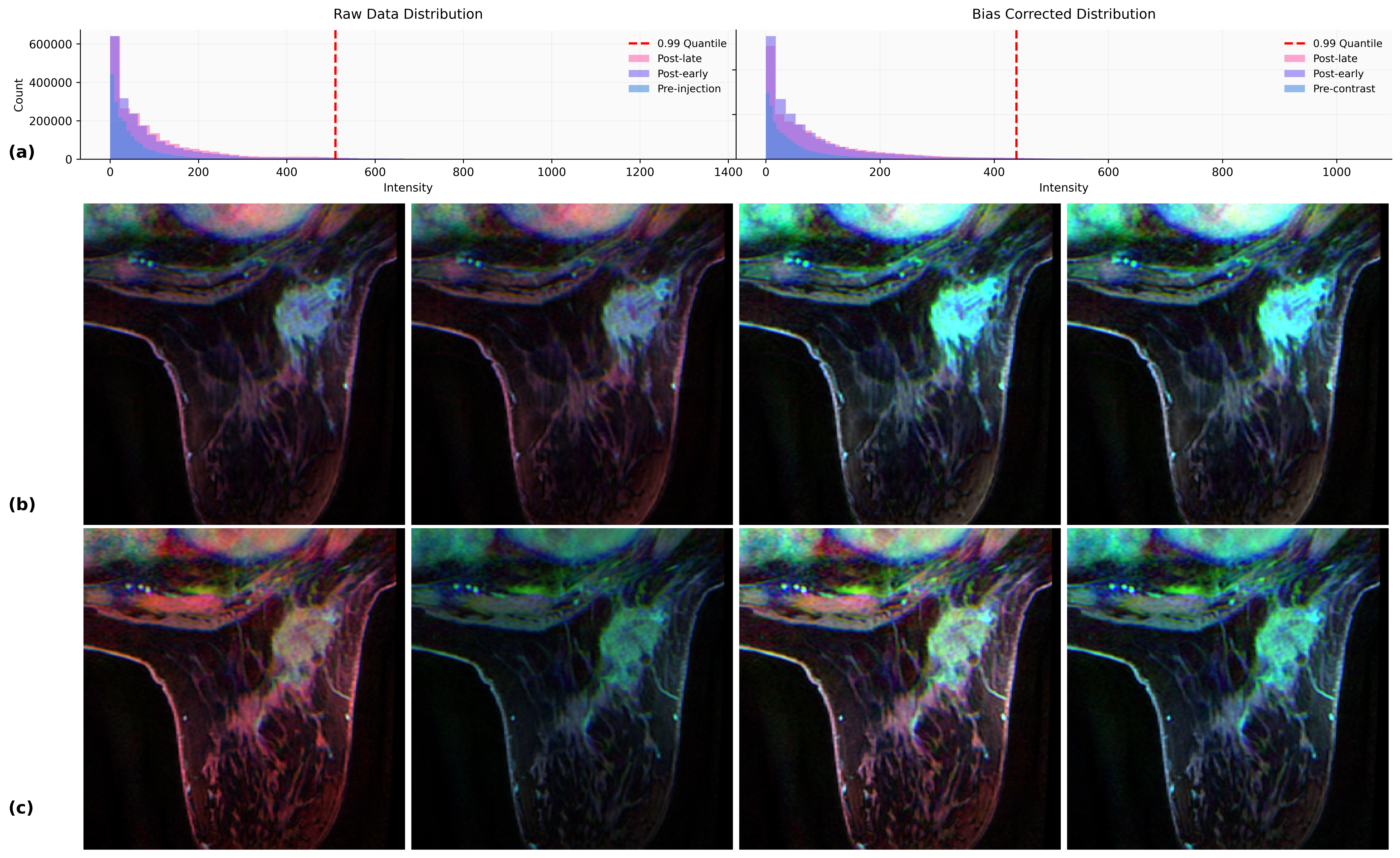}
\caption{Intensity normalization strategies for DCE-MRI (dynamic contrast-enhanced
magnetic resonance imaging) preprocessing. (a) Intensity distributions before (left) and after (right) bias field correction,
showing the three temporal acquisitions (pre-contrast, post-early, post-late) with the 0.99
quantile threshold marked. The threshold value is unitless, indicating a dimensionless
quantity. (b) Raw RGBfused images demonstrating four normalization approaches: global min-max
normalization, per-slice min-max normalization, $q_{0.95}$ upper clipping, and
$q_{0.99}$ upper clipping. (c) Corresponding bias-corrected images showing improved contrast consistency and reduced intensity variations across the four normalization strategies ($n=10$), including a scale bar representing 100~$\mu$m.}
\label{fig:preprocessing_norm}
\end{figure}

(Figure~\ref{fig:preprocessing_norm}) illustrates the impact of these strategies on intensity distributions and image contrast. Equation~\eqref{eq:slice_norm} through Equation~\eqref{eq:global_clip} define the transformations applied prior to model training.

\subsection{Preprocessing Strategy Evaluation for HER2 Prediction}

We developed a THDA-ResNet architecture for HER2 status prediction from
DCE-MRI and evaluated its performance across different preprocessing strategies. The challenge in preprocessing DCE-MRI for deep learning lies in converting 16-bit dynamic temporal sequences into standardized 8-bit RGB input while preserving diagnostically relevant contrast enhancement patterns. 

We systematically compared seven intensity normalization and clipping strategies: per-slice and global min-max normalization, quantile-based clipping, and global and channelwise upper clipping at multiple percentiles ($q_{0.99}$, $q_{0.95}$, $q_{0.98}$). Additionally, we tested N4 bias field correction variants for each method. N4 bias field correction mitigates low-frequency intensity inhomogeneities caused by magnetic field variations, effectively reducing the long tail in intensity distributions (Figure~\ref{fig:preprocessing_norm}a) and improving inter-scan comparability. All preprocessing approaches and their effects on intensity distributions and image appearance are visualized in (Figure~\ref{fig:preprocessing_norm}).

For comparative evaluation, we tested two transformer-based architectures: CvT~\cite{wu2021cvt} and ViT~\cite{dosovitskiy2020image}.
 (Tables~\ref{tab:her2_models}) and \ref{tab:her2_attresnet_upperclip} present HER2 prediction performance across models and preprocessing strategies.

\subsection{Deep Learning Modeling}

Our primary objective was to systematically compare the performance of
CNNs and transformer-based architectures for HER2 status prediction from DCE-MRI
scans. Following established methods~\cite{fridman2025breastdcedl}, RGB fusion was performed using three selected time points: pre-contrast, early post-contrast (peak enhancement), and late post-contrast (washout), as chosen by expert radiologists for their tumor segmentation analysis. 

\subsubsection{Triple-Head Dual-Attention ResNet Architecture}

\begin{figure}[htbp]
\centering
\includegraphics[width=\linewidth]{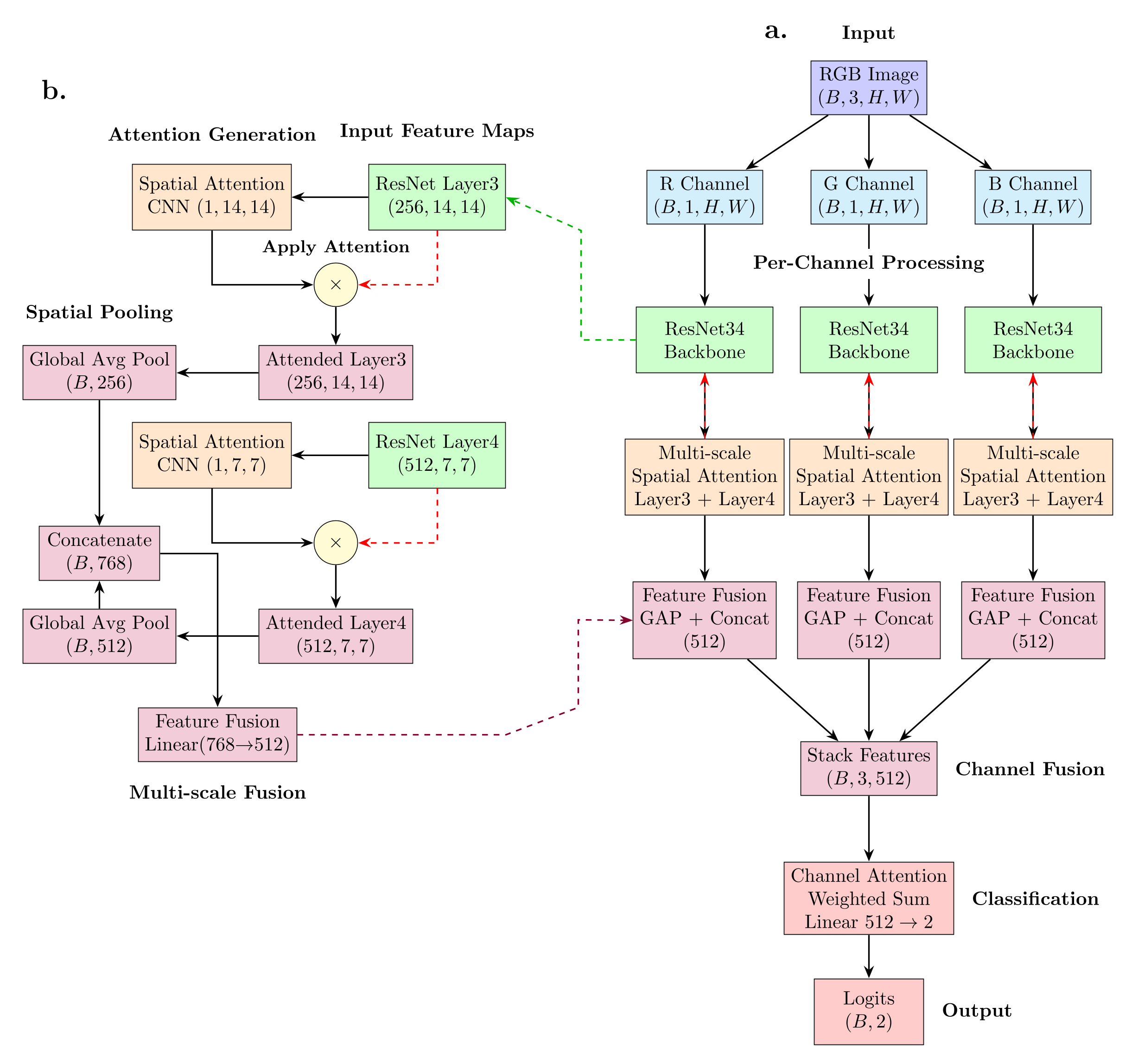}
\caption{Triple-Head Dual-Attention ResNet architecture for DCE-MRI analysis. (a) Overall architecture
showing per-channel processing of RGB-fused temporal phases through independent
ResNet34 backbones with multi-scale spatial attention mechanisms, followed by channel
fusion and classification. (b) Multi-scale spatial attention module detail demonstrating feature
extraction from ResNet Layer3 and Layer4, spatial attention generation, element-wise
multiplication for feature refinement, and multi-scale fusion through global average
pooling and concatenation. }
\label{fig:architecture}
\end{figure}

The proposed model, Triple-Head Dual-Attention ResNet (THDA-ResNet), is a convolutional neural network designed for quantitative analysis of breast DCE-MRI. As illustrated in (Figure~\ref{fig:architecture}), the network processes three temporal phases---pre-contrast, early post-contrast, and late post-contrast---by channel-wise routing each through independent, weight-shared ResNet34 backbones. Multi-scale spatial features are extracted at intermediate layers (Layer3 and Layer4), with each branch outputting feature maps of sizes $14 \times 14 \times 256$ and $7 \times 7 \times 512$, respectively.

To enhance spatial localization, specialized attention modules are
appended to both intermediate layer and Layer4 outputs.Layer4 outputs. These modules consist of stacked convolutional and batch normalization
layers followed by Rectified Linear Unit non-linearity, dropout, and sigmoid gating. The resulting spatial attention maps softly gate their corresponding feature maps through element-wise multiplication, isolating anatomically salient regions. An explicit edge suppression step down-weights border pixels to $30\%$ of their original value, mitigating border artifacts common in medical imaging.

Attention-refined multi-scale features are globally pooled and concatenated, capturing both fine and coarse spatial context. The resulting vectors from all three temporal branches are stacked and processed via an adaptive channel attention mechanism employing softmax operation, enabling dynamic weighting and aggregation of features from all temporal phases.

The fused feature vector undergoes layer normalization and dropout regularization prior to classification via a fully connected output layer. All attention module weights are initialized using Kaiming normalization~\cite{he2015delving}. The framework is implemented in PyTorch and trained end-to-end using Adaptive
Moment Estimation with Weight Decay optimizer and cross-entropy loss.

(Figure~\ref{fig:architecture}a) summarizes the overall triple-head processing pipeline. (Figure~\ref{fig:architecture}b) provides detailed view of the multi-scale spatial attention module, highlighting feature extraction, attention generation, spatial feature refinement, and fusion through global pooling and concatenation.

\begin{figure}[htbp]
\centering
\includegraphics[width=\linewidth]{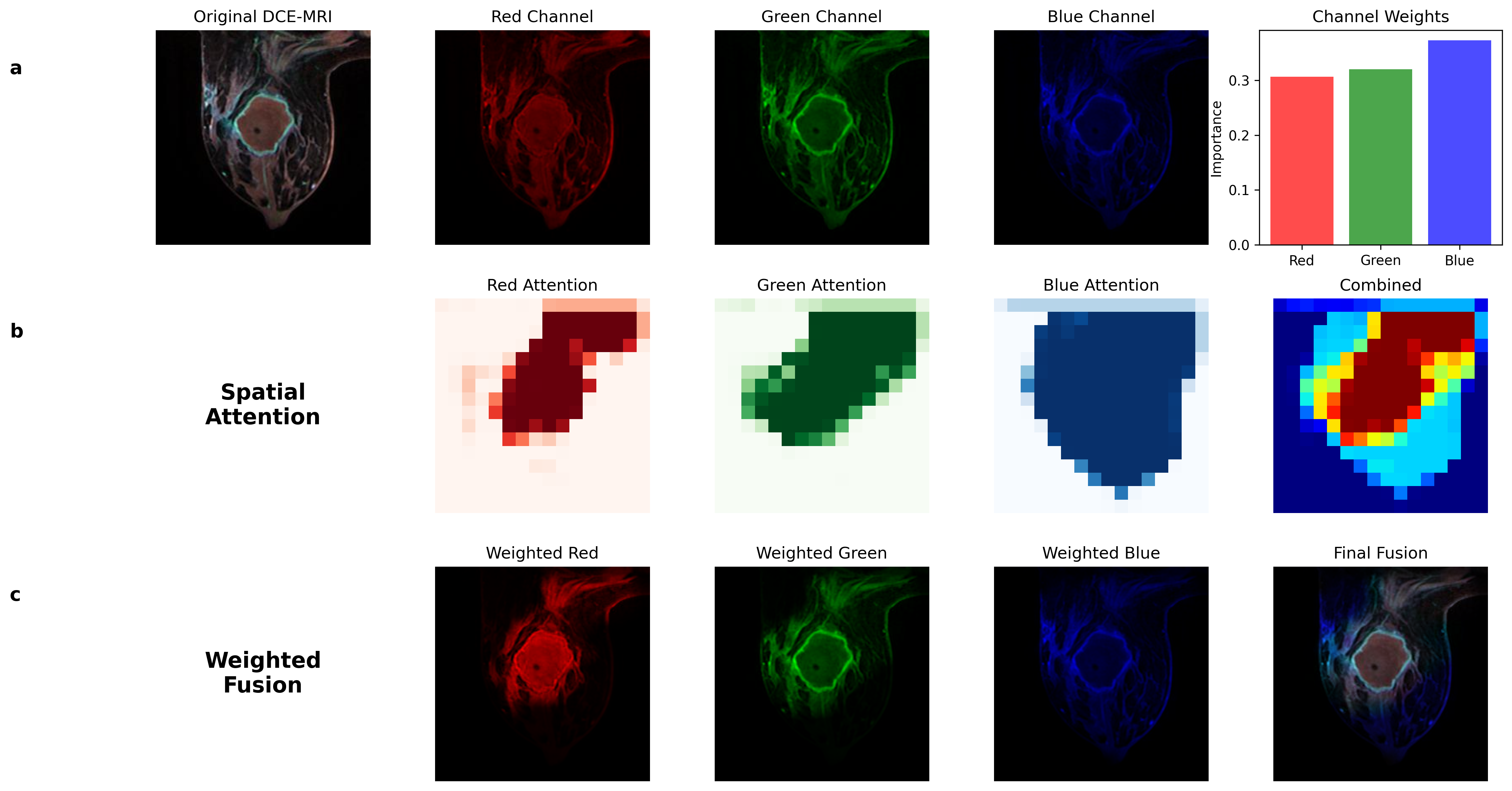}
\caption{Triple-Head Dual-Attention ResNet mechanism visualization for DCE-MRI (dynamic
contrast-enhanced magnetic resonance imaging) temporal fusion. (a) Input decomposition
showing the original RGB-fused DCE-MRI image and its constituent channels (red: precontrast, green: early post-contrast, blue: late post-contrast), with learned channel
attention weights indicating relative importance of each temporal phase. (b) Spatial
attention maps generated by the multi-scale attention module (Layer3, 14×14
resolution) for each temporal channel, highlighting tumor-relevant regions; combined
spatial attention demonstrates consensus across temporal phases. (c) Weighted fusion
process showing attention-modulated individual channels and their integration into the
final fused representation.}
\label{fig:combine_resatt}
\end{figure}

\subsubsection{Comparative Architectures}

For comparative evaluation, we tested two transformer-based architectures: Convolutional Vision Transformer (CvT)~\cite{wu2021cvt} and Vision Transformer (ViT)~\cite{dosovitskiy2020image}. Both models were pretrained on ImageNet and fine-tuned on our DCE-MRI dataset using identical preprocessing and training protocols as THDA-ResNet.

\subsection{Patient-Level Prediction Pipeline}

For each subject, several consecutive slices centered on the tumor were selected to comprehensively represent the lesion. From each slice, a $256 \times 256$ region was extracted, such that the crop was centered on the tumor and encompassed the entire bounding box, thereby aligning with the input size required by the network. These image patches were subjected to the preprocessing steps described previously and input into the trained neural network model. Prediction scores were generated for each slice; these slice-level predictions were then aggregated using summary statistics, specifically the mean, median, and minimum, to yield a robust patient-level prediction for downstream analysis.

We tested several aggregation methods to combine slice-level predictions into a patient-level score: mean \(\left(\frac{1}{N}\sum_i p_i\right)\), median, minimum, 10th (\(q_{0.1}\)) and 90th (\(q_{0.9}\)) percentiles, and the proportion of HER2-positive predictions. The median gave the best AUC, while the $90^\text{th}$ percentile provided the highest accuracy.

\subsection{External Data Validation}

To assess model generalizability, we performed external validation on the public BreastDCEDL\_AMBL dataset, which provides segmented tumor masks and HER2 status annotations for 42 patients~\cite{fridm2025ambl}. No training or fine-tuning was performed on this external dataset, ensuring that evaluation was strictly on unseen data. Input preparation and inference procedures mirrored those used during initial training, maintaining methodological consistency. For each patient, slice-level prediction scores were aggregated using summary statistics to produce final patient-level predictions. This approach demonstrates the clinical robustness of our model and its applicability to data collected from an independent institution with distinct imaging protocols~\cite{fridm2025ambl}.
\section{Results}

\subsection{Model Performance Comparison}

The THDA-ResNet architecture substantially outperformed transformer-based approaches for HER2 prediction at threshold 0.5 (Table~\ref{tab:her2_models}), THDA-ResNet with median aggregation achieved AUCs ranging from 0.72–0.74 across different preprocessing methods, with accuracy values of 0.50–0.57. In contrast, transformer models showed markedly lower performance: ViT with quantile aggregation achieved AUCs of 0.64–0.67 with accuracy of 0.67–0.74, while CvT with min aggregation reached AUCs of 0.61–0.63 with accuracy of 0.67–0.75. Despite transformers showing higher accuracy values, their substantially lower AUCs and sensitivity (0.19–0.47 for transformers vs 0.78–0.94 for THDA-ResNet) indicate they achieved specificity at the cost of missing positive cases.

For THDA-ResNet, global and channelwise upper clipping strategies ($q_{0.99}$, $q_{0.95}$, $q_{0.98}$) consistently yielded the best performance. N4 bias correction did not improve performance and was therefore excluded from transformer evaluations and subsequent analyses.



\begin{table*}[ht]
\caption{HER2 (Human Epidermal growth factor Receptor 2) prediction performance
across preprocessing methods at threshold 0.5.}
\begin{tabular}{lccccccccccc}
\toprule
\multicolumn{12}{c}{\textbf{Triple-Head Dual-Attention ResNet (Median aggregation)}} \\
Preprocessing                          & ACC   & AUC   & Sens  & Spec  & PPV   & NPV   & TP  & TN  & FP  & FN  \\
\midrule
Per-slice min-max normalization        & 0.568 & 0.723 & 0.781 & 0.500 & 0.333 & 0.877 & 25 & 50 & 50 & 7  \\
Global min-max normalization           & 0.462 & 0.700 & 0.906 & 0.320 & 0.299 & 0.914 & 29 & 32 & 68 & 3  \\
Lower quantile clipping ($q_{0.1}$)    & 0.568 & 0.720 & 0.781 & 0.500 & 0.333 & 0.877 & 25 & 50 & 50 & 7  \\
Global upper clipping ($q_{0.99}$)     & 0.500 & 0.726 & \textbf{0.938} & 0.360 & 0.319 & \textbf{0.947} & 30 & 36 & 64 & 2  \\
Channel upper clipping ($q_{0.99}$)    & 0.515 & 0.731 & 0.844 & 0.410 & 0.314 & 0.891 & 27 & 41 & 59 & 5  \\
Channel upper clipping ($q_{0.95}$)    & 0.455 & \textbf{0.734} & \textbf{0.938} & 0.300 & 0.300 & 0.938 & 30 & 30 & 70 & 2 \\
Channel upper clipping ($q_{0.98}$)    & 0.523 & \textbf{0.744} & 0.875 & 0.410 & 0.322 & 0.911 & 28 & 41 & 59 & 4  \\
N4 Per-slice min-max      & 0.568 & 0.719 & 0.781 & 0.500 & 0.333 & 0.877 & 25 & 50 & 50 & 7  \\
N4 Global min-max normalization        & 0.485 & 0.710 & \textbf{0.938} & 0.340 & 0.312 & 0.944 & 30 & 34 & 66 & 2  \\
N4 Lower quantile clipping ($q_{0.1}$) & \textbf{0.576} & 0.722 & 0.781 & \textbf{0.510} & \textbf{0.338} & 0.879 & 25 & 51 & 49 & 7  \\
N4 Global upper clipping ($q_{0.99}$)  & 0.500 & 0.687 & 0.906 & 0.370 & 0.315 & 0.925 & 29 & 37 & 63 & 3  \\
N4 Channel upper clipping ($q_{0.99}$) & 0.508 & 0.698 & 0.906 & 0.380 & 0.319 & 0.927 & 29 & 38 & 62 & 3  \\
N4 Channel upper clipping ($q_{0.95}$) & 0.394 & 0.686 & \textbf{0.938} & 0.220 & 0.278 & 0.917 & 30 & 22 & 78 & 2  \\
N4 Channel upper clipping ($q_{0.98}$) & 0.523 & 0.712 & 0.875 & 0.410 & 0.322 & 0.911 & 28 & 41 & 59 & 4  \\
\midrule
\multicolumn{12}{c}{\textbf{ViT Transformer (Quantile aggregation, no N4)}} \\
Preprocessing                          & ACC   & AUC   & Sens  & Spec  & PPV   & NPV   & TP  & TN  & FP  & FN  \\
\midrule
Per-slice min-max normalization        & 0.727 & \textbf{0.668} & 0.344 & \textbf{0.850} & 0.423 & 0.802 & 11 & 85 & 15 & 21  \\
Global min-max normalization           & \textbf{0.735} & 0.645 & \textbf{0.438} & 0.830 & \textbf{0.452} & \textbf{0.822} & 14 & 83 & 17 & 18  \\
Lower quantile clipping ($q_{0.1}$)    & \textbf{0.735} & \textbf{0.666} & 0.375 & \textbf{0.850} & \textbf{0.444} & 0.810 & 12 & 85 & 15 & 20  \\
Global upper clipping ($q_{0.99}$)     & 0.705 & 0.645 & 0.312 & 0.830 & 0.370 & 0.790 & 10 & 83 & 17 & 22  \\
Channel upper clipping ($q_{0.99}$)    & 0.674 & 0.646 & 0.312 & 0.790 & 0.323 & 0.782 & 10 & 79 & 21 & 22  \\
Channel upper clipping ($q_{0.95}$)    & 0.682 & 0.642 & 0.250 & 0.820 & 0.308 & 0.774 &  8 & 82 & 18 & 24  \\
Channel upper clipping ($q_{0.98}$)    & 0.697 & 0.644 & 0.312 & 0.820 & 0.357 & 0.788 & 10 & 82 & 18 & 22  \\
\midrule
\multicolumn{12}{c}{\textbf{CvT Transformer (Min aggregation, no N4)}} \\
Preprocessing                          & ACC   & AUC   & Sens  & Spec  & PPV   & NPV   & TP  & TN  & FP  & FN  \\
\midrule
Per-slice min-max normalization        & 0.727 & \textbf{0.625} & 0.312 & 0.860 & 0.417 & 0.796 & 10 & 86 & 14 & 22  \\
Global min-max normalization           & 0.697 & 0.614 & \textbf{0.469} & 0.770 & 0.395 & \textbf{0.819} & 15 & 77 & 23 & 17 \\
Lower quantile clipping ($q_{0.1}$)    & 0.720 & 0.622 & 0.312 & 0.850 & 0.400 & 0.794 & 10 & 85 & 15 & 22  \\
Global upper clipping ($q_{0.99}$)     & 0.667 & 0.617 & 0.375 & 0.760 & 0.333 & 0.792 & 12 & 76 & 24 & 20  \\
Channel upper clipping ($q_{0.99}$)    & 0.720 & 0.607 & 0.188 & 0.890 & 0.353 & 0.774 &  6 & 89 & 11 & 26  \\
Channel upper clipping ($q_{0.95}$)    & \textbf{0.750} & \textbf{0.627} & 0.219 & \textbf{0.920} & \textbf{0.467} & 0.786 &  7 & 92 & 8 & 25  \\
Channel upper clipping ($q_{0.98}$)    & 0.705 & 0.607 & 0.312 & 0.830 & 0.370 & 0.790 & 10 & 83 & 17 & 22  \\
\bottomrule
\end{tabular}
\label{tab:her2_models}

\begin{minipage}{0.97\linewidth}
\small
\textbf{Notes for (Table~\ref{tab:her2_models}:})
\begin{itemize}
  \item \textbf{ACC} – Accuracy: proportion of correct predictions.
  \item \textbf{AUC} – Area under the ROC curve: overall discriminative performance.
  \item \textbf{Sens} – Sensitivity: true positive rate = TP / (TP + FN).
  \item \textbf{Spec} – Specificity: true negative rate = TN / (TN + FP).
  \item \textbf{PPV} – Positive predictive value (precision): TP / (TP + FP).
  \item \textbf{NPV} – Negative predictive value: TN / (TN + FN).
  \item \textbf{TP} – True positives.
  \item \textbf{TN} – True negatives.
  \item \textbf{FP} – False positives.
  \item \textbf{FN} – False negatives.
  \item Best results per model are shown in \textbf{bold}.
\end{itemize}
\end{minipage}
\end{table*}

\subsection{Optimal THDA-ResNet Configuration}

(Table~\ref{tab:her2_attresnet_upperclip}) presents the top-performing THDA-ResNet configurations using median aggregation with upper clipping normalization (no N4 correction) at threshold 0.7. This higher threshold improved the specificity-sensitivity balance compared to threshold 0.5. The three best preprocessing methods—channel upper clipping $q_{0.98}$, channel upper clipping $q_{0.95}$, and global upper clipping $q_{0.99}$—achieved accuracy of 0.74–0.75 with AUCs of 0.72–0.74 on the complete test cohort. Performance on the ISPY2 subset was slightly higher, with accuracy of 0.75–0.77 and AUCs of 0.72–0.75.

Channel upper clipping at $q_{0.98}$ yielded the best overall performance: accuracy 0.75, AUC 0.74, sensitivity 0.41, and specificity 0.86 on the full test set. The higher threshold (0.7 vs 0.5) substantially improved specificity (0.84–0.87 vs 0.30–0.51) while maintaining reasonable sensitivity (0.31–0.41 vs 0.78–0.94), resulting in better-balanced predictions suitable for clinical decision support.

\begin{table}[ht]
\caption{Top THDA-ResNet (Threshold-based Deep Hashing Association Network)
configurations at threshold 0.7 (median aggregation, upper clipping, no N4).}
\begin{tabular}{lccccccccccc}
\toprule
\multicolumn{12}{c}{\textbf{Complete Test Cohort - 132 cases}} \\
Preprocessing                          & ACC   & AUC   & Sens  & Spec  & PPV   & NPV   & TP  & TN  & FP  & FN  \\
\midrule
Channel upper clipping ($q_{0.98}$)    & \textbf{0.750} & \textbf{0.744} & \textbf{0.406} & 0.860 & \textbf{0.481} & \textbf{0.819} & 13 & 86 & 14 & 19  \\
Channel upper clipping ($q_{0.95}$)    & 0.737 & 0.721 & 0.364 & 0.844 & 0.400 & 0.823 &  8 & 65 & 12 & 14  \\
Global upper clipping ($q_{0.99}$)     & 0.735 & 0.731 & 0.312 & 0.870 & 0.435 & 0.798 & 10 & 87 & 13 & 22  \\
\midrule
\multicolumn{12}{c}{\textbf{ISPY2 Dataset Test cohort - 99 cases}} \\
Preprocessing                          & ACC   & AUC   & Sens  & Spec  & PPV   & NPV   & TP  & TN  & FP  & FN  \\
\midrule
Channel upper clipping ($q_{0.98}$)    & \textbf{0.768} & \textbf{0.753} & \textbf{0.438} & \textbf{0.875} & \textbf{0.500} & \textbf{0.840} & 14 & 62 & 9 & 18  \\
Channel upper clipping ($q_{0.95}$)    & 0.758 & 0.735 & 0.375 & 0.863 & 0.462 & 0.829 & 12 & 61 & 10 & 20  \\
Global upper clipping ($q_{0.99}$)     & 0.748 & 0.724 & 0.344 & 0.887 & 0.478 & 0.821 & 11 & 63 & 8 & 21  \\
\bottomrule
\end{tabular}
\label{tab:her2_attresnet_upperclip}

\vspace{1ex}
\begin{minipage}{\textwidth}
\small
\textbf{Notes for (Table~\ref{tab:her2_attresnet_upperclip}:})
\begin{itemize}
  \item Metric definitions are consistent with those in (Table~\ref{tab:her2_models}).
  
\end{itemize}
\end{minipage}
\end{table}
\subsection{External Validation on BreastDCEDL\_AMBL Dataset}

To assess model generalizability beyond the I-SPY trials, we performed external validation on the public BreastDCEDL\_AMBL dataset~\cite{fridm2025ambl}, which provides segmented tumor masks and HER2 status annotations for 43 lesions from an independent institution. No training, fine-tuning, or hyperparameter adjustment was performed on this external dataset, ensuring strictly unseen data evaluation. The same THDA-ResNet model trained on I-SPY data was applied directly with median aggregation at threshold 0.5.

(Table~\ref{tab:ambl_validation}) presents external validation performance across the same preprocessing strategies evaluated on I-SPY. Channel upper clipping at $q_{0.98}$ achieved the best overall performance with accuracy 0.65, AUC 0.61, and balanced sensitivity (0.69) and specificity (0.63). Global min-max normalization yielded the highest accuracy (0.67) and AUC (0.66), though with lower sensitivity (0.56) compared to upper clipping methods.

Notably, the performance on this external dataset was lower than on I-SPY test data (AUC 0.61--0.66 vs 0.72--0.74), reflecting expected domain shift between institutions with different imaging protocols, patient populations, and scanner characteristics. Despite this gap, the model maintained discriminative ability (AUC >0.60 for most methods), demonstrating reasonable generalizability. The relative ranking of preprocessing strategies remained partially consistent, with upper clipping methods and min-max normalization performing competitively, supporting the robustness of these approaches across datasets.

\begin{table}[ht]
\centering
\caption{External validation on BreastDCEDL\_AMBL dataset (43 lesions) using THDA-ResNet trained on I-SPY data.}
\begin{tabular}{lccccccccccc}
\toprule
Preprocessing                          & ACC   & AUC   & Sens  & Spec  & PPV   & NPV   & TP  & TN  & FP  & FN & N \\
\midrule
Per-slice min-max normalization        & 0.651 & 0.655 & 0.500 & 0.741 & 0.533 & 0.714 & 8  & 20 & 7  & 8  & 43 \\
Global min-max normalization           & \textbf{0.674} & \textbf{0.664} & 0.562 & \textbf{0.741} & 0.562 & \textbf{0.741} & 9  & 20 & 7  & 7  & 43 \\
Lower quantile clipping ($q_{0.1}$)    & 0.651 & 0.655 & 0.500 & \textbf{0.741} & 0.533 & 0.714 & 8  & 20 & 7  & 8  & 43 \\
Global upper clipping ($q_{0.99}$)     & 0.628 & 0.616 & \textbf{0.688} & 0.593 & 0.500 & 0.762 & 11 & 16 & 11 & 5  & 43 \\
Channel upper clipping ($q_{0.99}$)    & 0.558 & 0.606 & 0.625 & 0.519 & 0.435 & 0.700 & 10 & 14 & 13 & 6  & 43 \\
Channel upper clipping ($q_{0.95}$)    & 0.535 & 0.597 & 0.625 & 0.481 & 0.417 & 0.684 & 10 & 13 & 14 & 6  & 43 \\
Channel upper clipping ($q_{0.98}$)    & 0.651 & 0.613 & \textbf{0.688} & 0.630 & \textbf{0.524} & \textbf{0.773} & 11 & 17 & 10 & 5 & 43 \\
\bottomrule
\end{tabular}
\label{tab:ambl_validation}

\vspace{1ex}
\begin{minipage}{\textwidth}
\small
\textbf{Notes for (Table~\ref{tab:ambl_validation}:})
\begin{itemize}
  \item Metric definitions are consistent with those in (Table~\ref{tab:her2_models}).
  \item N: total number of lesions in the external validation cohort.
  \item No training or fine-tuning was performed on this dataset.
  \item Best results shown in \textbf{bold}.
\end{itemize}
\end{minipage}
\end{table}

\section{Discussion}

We systematically evaluated preprocessing strategies and deep learning architectures for HER2 prediction from DCE-MRI using a multicenter cohort (n=1,149). Triple-Head Dual-Attention ResNet outperformed transformer-based models, with preprocessing strategy selection critically impacting performance across all architectures.

Different normalization strategies present distinct tradeoffs. Per-slice normalization maximizes local contrast within each slice but corrupts spatial intensity relationships across slices. Per-channel normalization (normalizing each temporal phase independently) maximizes contrast within each time point but corrupts temporal intensity relationships. Global normalization across all slices and time points preserves both spatial and temporal relationships but severely compresses dynamic range due to the long tail in MRI intensity distributions, where maximum values often exceed 12,000 while the 8-bit target range is only 0--255.

Our results demonstrate task-specific optimal strategies. Per-channel normalization yielded superior HER2 prediction performance, whereas previous work on pathologic complete response (pCR) prediction found global normalization performed better~\cite{fridman2025breastdcedl}. This suggests that HER2 status is encoded primarily in spatial morphological features within individual time points, while pCR response correlates with temporal enhancement kinetics reflecting vascularity and treatment response~\cite{kuhl2007current,wolff2018her2,song2021texture},  while therapy response depends on vascular characteristics and enhancement kinetics captured through temporal contrast dynamics~\cite{li2020predicting,kuhl2007current,song2021texture}.

Upper percentile clipping improved performance across architectures. However, optimal strategies varied between I-SPY and AMBL datasets, indicating that intensity normalization for deep learning requires further investigation to achieve robust cross-institutional generalization.

Notably, N4 bias field correction—standard in radiomics~\cite{tustison2010n4itk}—slightly degraded deep learning performance across nearly all experiments. This finding eliminates an unnecessary computationally expensive preprocessing step, demonstrating that radiomics preprocessing strategies may not transfer to deep learning~\cite{kubassova2007quantitative}.

Triple-Head Dual-Attention ResNet substantially outperformed transformers (AUC 0.74 vs 0.61--0.67), though this finding requires careful interpretation. The performance difference may lie primarily in how temporal information is presented to the model rather than inherent architectural superiority. Our ResNet architecture processes each temporal phase through separate heads before fusion, explicitly preserving phase-specific features, whereas transformers receive RGB-fused channels as direct input. This difference in data presentation may disadvantage transformers, which could potentially benefit from alternative tokenization strategies that maintain temporal separation before integration. Additionally, transformers typically require larger datasets~\cite{dosovitskiy2020image}, and our 885 training cases may be insufficient. Hybrid CNN-transformer architectures~\cite{wu2021cvt} may offer advantages for medical imaging with limited data.

Our systematic preprocessing benchmark demonstrated that multiple normalization strategies serve as effective augmentation for limited medical datasets. External validation (AUC 0.61--0.66 on 43 lesions) demonstrated reasonable cross-institutional generalizability despite substantial domain shift from different scanners, protocols, and populations, showing the model's potential to generalize when sufficient training data is available.

Several limitations warrant discussion. Our 2D approach does not exploit 3D structure. We focused exclusively on DCE sequences, whereas multiparametric models incorporating T2-weighted and diffusion-weighted imaging could improve performance. We evaluated binary classification, though HER2 expression exists on a spectrum, and ordinal classification approaches could better handle equivocal cases. Our optimal configuration (accuracy 0.75, specificity 0.86) is insufficient for standalone diagnosis. Beyond performance metrics, understanding which spatial features and patterns drive model predictions through attention map analysis could inform radiological interpretation and guide future biomarker development, though validation is needed to determine whether these learned representations correspond to known biological characteristics.

Future work should investigate alternative temporal modeling and data presentation strategies for transformers, explore multiparametric integration including T2-weighted and diffusion-weighted sequences, and expand public datasets with diverse institutions and comprehensive acquisition protocols. Critically, research is needed to validate the clinical interpretability and utility of attention-based features for medical decision support.

This study demonstrates that per-channel normalization and upper percentile clipping optimize HER2 prediction from DCE-MRI, with Triple-Head Dual- Attention ResNet achieving AUC 0.74 and reasonable cross-institutional generalizability (AUC 0.61--0.66). Notably, N4 bias correction—standard in radiomics—degraded deep learning performance, suggesting that conventional preprocessing strategies may not transfer to learned representations. These findings advance reproducible deep learning approaches for breast cancer biomarker prediction from medical imaging.


\section{Ethics Statement}

This study used publicly available deidentified datasets from The Cancer Imaging Archive (TCIA), including I-SPY 1, I-SPY 2, and Duke Breast Cancer MRI datasets. All original data collection for these datasets was conducted with appropriate institutional review board (IRB) approvals at the respective participating institutions, with informed consent obtained from all participants as described in the original publications. No additional ethical approval was required for this secondary analysis of publicly available anonymized data. All data handling and analysis procedures were in accordance with the relevant data protection regulations and institutional guidelines of Ariel University.

\section{Conflicts of Interest}
The authors declare no conflicts of interest related to this work.
\FloatBarrier

\bibliographystyle{unsrt}
\bibliography{ref}  

\end{document}